\documentstyle[12pt]{article}
\def\beq{\begin{equation}}
\def\eeq{\end{equation}}
\def\bea{\begin{eqnarray}}
\def\eea{\end{eqnarray}}

\def\ba{\begin{array}}
\def\ea{\end{array}}
\def\bce{\begin{center}}
\def\ece{\end{center}}

\begin{document}
\begin{titlepage}
\rightline{BSRI-POSTECH-9704001}
\def\today{\ifcase\month\or
        January\or February\or March\or April\or May\or June\or
        July\or August\or September\or October\or November\or December\fi,
  \number\year}
\rightline{hep-th/9704061}
\vskip 1cm
\centerline{\Large \bf Geometry, D-Branes and  $N=1$ Duality in 
Four Dimensions I}
\vskip 1cm
\centerline{\sc Changhyun Ahn \footnote{chahn@nms.kyunghee.ac.kr}}
\centerline {{\it Department of Physics and}}
\centerline{{\it Research Institute for Basic Sciences,}} 
\centerline {{\it Kyung Hee University,}}
\centerline {{\it Seoul 130-701, Korea}}
\centerline {and}
\centerline{\sc Kyungho Oh \footnote{oh@arch.umsl.edu, 
oh@vision.postech.ac.kr}}
\centerline{{ \it Department of Mathematics,}}
\centerline{{\it University of Missouri-St. Louis,}}
\centerline{{\it St. Louis, Missouri 63121, USA}}
\vskip 1cm
\centerline{\sc Abstract}
\vskip 0.1in
We consider $N=1$ dualities in four dimensional supersymmetric gauge theories
as a geometrical realization of wrapping D 6-branes around 3-cycles of
Calabi-Yau threefolds in type IIA string theory. 
By extending the recent work of Ooguri and Vafa to the case of
$SU, SO$ and $Sp$ gauge groups with  {\it additional} fields together
with defining fields, 
we give simple geometrical descriptions of the interrelation 
between the electric theory and its magnetic dual in terms of the
configuration of D 6-branes wrapped 3-cycles.
\vskip 0.3in
\leftline{04/01/97}
\end{titlepage}
\newpage

\section{ Introduction}
\setcounter{equation}{0}

String theory interprets many nontrivial aspects of four dimensional
supersymmetric field theory by exploiting T-duality on the
local model for the compactification manifold.
The compactification of F-theory (defined as type IIB string compactification
where the dilaton and the axion are not constant but vary over the internal
manifold) on elliptic Calabi-Yau(CY) fourfolds from 12 dimensions leads to
$N=1$ supersymmetric field theories in four dimensions.
For the case of pure $SU(N_c)$ Yang-Mills gauge theory, 
it has been studied in \cite{KV1}
that the gauge symmetry can be obtained in terms of the structure of
the D(irichlet) 7-brane worldvolume, that is the product of uncompactified 
four
dimensional spacetime and four dimensional subspace of base threefold 
of CY fourfolds,  
over which the elliptic fibration has ADE type of 
singularity. 
By adding some numbers of D 3-branes and bringing them
near the complex 2-dimensional surface (which is the 
compact part of D 7-brane worldvolume), it has been shown in 
\cite{BJPSV} that
the local string model gives rise to matter hypermultiplets in the 
fundamental representation with
pure $SU(N_c)$ Yang-Mills theory. Moreover, Seiberg's duality \cite{Seiberg} 
for the $N=1$
supersymmetric field theory can be mapped to T-duality exchanging the
D 3-brane charge and D 7-brane charge even though we are working on F-theory
which has no T-duality generically. However this makes sense in the fact that
F-theory local model can be realized by D 7-branes of type IIB string theory
for which the T-duality symmetry holds. 
It has been further discussed in \cite{VZ}
that for the extension of $SO(N_c)$ and $Sp(N_c)$ 
gauge theories \cite{IS,IP} coupled
to matter the local string models are type IIB orientifolds, for which 
T-duality symmetry applies, with D 7-branes on a curved orientifold
7-plane.

On the other hand, 
{\it other} simple interpretation of $N=1$ duality for $SU(N_c)$ gauge theory
was given in \cite{EGK} by
D-brane description in the presence of
NS 5-branes in a flat geometry according to the approach of \cite{HW}. 
Extension of this to the case of $SO(N_c)$ and $Sp(N_c)$
gauge theories with flavors was presented in \cite{EJS} by
inserting an orientifold 4-plane (See also the relevant papers appeared 
in \cite{BH,BSTY} along the line of this approach).

Recently, Ooguri and Vafa have considered in \cite{OV} that
$N=1$ duality can be embedded into 
type IIA string theory with D 6-branes, partially
wrapped around three cycles of CY threefold, filling four
dimensional spacetime.
They discussed what happens to the wrapped cycles and studied
the relevant field theory results when a transition in  the moduli of CY 
threefolds occurs. Furthermore, they reinterpreted the configuration of
D-branes in the presence of NS 5-branes \cite{EGK} 
as purely classical geometrical
realization.

In this paper, we generalize the approach of \cite{OV} to various models,
consisting of {\it one or two 2-index tensors} 
and some fields in the defining
representation (fundamental representation for $SU(N_c)$ and $Sp(N_c)$,
vector representation for $SO(N_c)$), presented earlier by many 
authors \cite{Kutasov,Intril,LS,ILS} and study its geometric realizations
by wrapping D 6-branes about 3-cycles of CY threefolds. 

\section{Geometrical Realization of $N=1$ Duality}
\setcounter{equation}{0}

Let us start with the compactification of type IIB string theory
on the CY threefold leading to $N=2$ supersymmetric field theories in 4 
dimensions. Suppose a number of D 3-branes wrapped around
a set of three cycles of CY threefold. It is 
known \cite{CDFV} that whenever
the integration of the holomorphic 3-form on the CY threefold 
around three cycles  
form  parallel vectors in the complex plane, such a D 3-brane configuration
give rises to a BPS state.
Therefore we obtain type IIA string theory with D 6-branes, partially
wrapped around three cycles of CY threefold,
filling 4 dimensional spacetime
after T-dualizing the 3-spatial directions of three torus
$T^3$. We end up with 
$N=1$ supersymmetric field theories in 4 dimensions from the BPS states
of $N=2$ string theory. 
The local model of CY threefold can be described by \cite{BSV,OV}
five complex coordinates $x, y, x', y'$ and $z$ satisfying the following
equations:
$$
x^2+y^2=\prod_i (z-a_i), \;\;\;\; x'^2+y'^2=\prod_j (z-b_j)
$$ 
where each of $C^*$'s is embedded in $(x, y)$-space and $(x', y')$-space
respectively over a generic point $z$. 
More precisely, this describes a family of a product of
two copies of one-sheeted hyperboloids in $(x, y)$-space 
and $(x', y')$-space respectively parameterized
by the $z$-coordinates. 
For a fixed $z$ away from $a_i$ and $b_j$ there
exist nontrivial $S^1$'s in each of $C^*$'s corresponding to the waist
of the hyperboloids. It is easy to see that
when $z=a_i$ or $z=b_j$ the corresponding circles vanish as the waists shrink.
Then we regard 3 cycles as the product of $S^1 \times S^1$ cycles over
each point on the $z$-plane, with the segments in the $z$-plane ending
on $a_i$ or $b_j$. When we go between two $a_i$'s ($b_j$'s) 
without passing through
$b_j$ ($a_i$) the 3 cycles sweep out $S^2 \times S^1$.
On the other hand, when we go between $a_i$ and $b_j$ the 3 cycle becomes
$S^3$. We will denote the 3-cycle lying over between $a_i$ and $b_j$
by $[a_i, b_j]$ and also denote other cycles in a similar fashion.

From now on we consider particular $N=1$ supersymmetric field theories
and see how their $N=1$ dualities arise from the  transition in  moduli
space of CY threefolds.
  
{\bf 1) $SU(N_c)$ with an adjoint field and $N_f$ fundamental 
flavors\cite{Kutasov,ILS}:}

We study supersymmetric Yang-Mills theory with gauge group $SU(N_c)$,
a chiral matter superfield $X$ in the adjoint representation of $SU(N_c)$,
$N_f$ fundamental multiplets $Q^i$, and $N_f$ antifundamental
multiplets $\widetilde{{Q}_{\widetilde{i}}}$ where $i, 
\widetilde{i}=1, \cdots, N_f$. The superpotential is
$\mbox{Tr} X^{k+1}$. 
The theory given by this superpotential has a stable vacuum if and only
if $k N_f \geq N_c$.
For $k=1$, the dual theory, which has gauge group
$SU(N_f-N_c)$ with adjoint superfield $Y$, $N_f$ flavors of
magnetic quarks $q^i, \widetilde{q_{\widetilde{i}}}$ and magnetic meson
since $X$ and $Y$ are massive and can be integrated out, 
was realized geometrically
in \cite{OV}
by the configuration of points ordered as $(b, a_1, a_2)$ along the real
axis with $(N_f-N_c)$
 D-branes\footnote{We denote D 6-branes by D branes for simplicity.} 
wrapping around $[b, a_1]$ cycle and $N_f$ D-branes wrapping
around $[a_1, a_2]$ cycle. Recall that the electric theory corresponds to
the configuration of $N_c$ D-branes around the cycle $[a_1, b]$
and $N_f$ D-branes around $[b, a_2]$ cycle after moving $a_1$ to the 
left side of $b$.

We now proceed the case of $k=2$ and discuss how the $N=1$ duality is
realized geometrically by D-brane picture.
Let us consider that there are two points $a_1$ and $a_2$ along the real
part of $z$-plane where the first $C^*$ degenerates and two
points $b_1$ and $b_2$ along the real axis between $a_1$ and $a_2$ where
the second $C^*$ degenerates. Suppose we have four ordered special
points $(a_1, b_1, b_2, a_2)$ along the real axis. 
Then the three cycle $[a_1, b_1]$ lying between $a_1$ and $b_1$ is $S^3$ 
and the three cycle $[b_1, b_2]$ lying between $b_1$ and $b_2$ is
 $S^1 \times S^2$. Thus the three cycle $[a_1, b_2]$ is a bouquet of $S^3$
and $S^1\times S^2$ joined together at $z=b_1$.
Then we wrap $N_1$ D-branes around the
three cycle $[a_1, b_1]$ and $N_2$ D-branes around
 three  cycle
$[a_1, b_2]$ such that $N_1+N_2=N_c$ (because in the limit of
$b_1 \rightarrow b_2$ this system should be consistent with 
$N_c$ D-branes around the cycle $[a_1, b_2]$) 
and $N_f$ D-branes around the three  cycle $[b_1, a_2]$ and
$N_f$ D-branes around the three  cycle $[b_2, a_2]$ where we assume 
$2 N_f \geq N_c$. 

Now we want to move to other point in
 the moduli of CY threefolds and end up with
the configuration in which the degeneration points are along the real
$z$-axis except that the orders are changed from $(a_1, b_1, b_2, a_2)$
to $(b_1, b_2, a_1, a_2)$. As done in \cite{OV}, we push the point $b_1$
up along the imaginary direction since we have the freedom to
turn on a Fayet-Iliopoulos(FI) D term. Then $(N_1+N_2)$ of D-branes connect
directly between $(a_1, b_2)$ and $(N_2+N_f-N_1-N_2)$ of D-branes go
between $(b_1, b_2)$. We continue to move $b_1$ along the negative
real axis and pass the $x$-coordinate of $a_1$ and push down it to the real
axis. At this moment, the $(N_f-N_1)$ D-branes which were between 
$(b_1, b_2)$ decompose to $(N_f-N_1)$ D-branes between $(b_1, a_1)$ and
$(N_f-N_1)$ D-branes between $(a_1, b_2)$ which amounts to the decomposition
of the three cycle $[b_1, b_2]$ into a bouquet of
two 3-cycles of $S^3$. 
The $(N_1+N_2)$ D-branes which were
going between $(a_1, b_2)$ will recombine with the newly generated
$(N_f-N_1)$ D-branes
ending up with the total of $(N_f+N_2)$ D-branes along $[a_1, b_2]$ cycle.
Similarly we do push the point $b_2$ in turn and move it between
$b_1$ and $a_1$ using the above operation.  
Then $(N_f+N_2)$ of D-branes connect
directly between $(a_1, a_2)$ and $(2 N_f-N_2- N_f)$ of D-branes go
between $(b_2, a_2)$. 
We can see that the $(N_f-N_2)$ D-branes which were between 
$(b_2, a_2)$ decompose to $(N_f-N_2)$ D-branes between $(b_2, a_1)$ and
$(N_f-N_2)$ D-branes between $(a_1, a_2)$.
The $(N_f+N_2)$ D-branes which were
going between $(a_1, a_2)$ will recombine with the new $(N_f-N_2)$ D-branes
ending up with the total of $2 N_f$ D-branes along $[a_1, a_2]$ cycle.
Therefore the final configuration is a configuration of points ordered
as $(b_1, b_2, a_1, a_2)$ with $(N_f-N_1)$ D-branes wrapped around
$[b_1, b_2]$ and $(2 N_f-N_1-N_2)$ D-branes wrapped around $[b_2, a_1]$
and $2 N_f$ D-branes wrapped around $[a_1, a_2]$.
Notice that the number of D-branes along the cycle $[b_2, a_2]$ in the
original configuration are the
same of those along the cycle $[a_1, a_2]$ after we moved the points
$b_1$ and $b_2$.
In the limit 
$b_1 \rightarrow b_2$, we note that this is the magnetic description
of the original theory. The gauge 
group \cite{Kutasov} is $SU(\widetilde{N_c})=SU(2 N_f-N_c)$ due to the fact
that $N_1+N_2=N_c$.
In addition to the dual quarks $q^i, \widetilde{q_{\widetilde{i}}}$ 
and adjoint field $Y$
we have two singlet chiral superfields $M_1, M_2$ which interact with
the dual quarks through the superpotential in the magnetic theory.

We expect that for general value of $k$, the above procedure can be
done similarly. Suppose
 we wrap $N_1$ D-branes around the
three cycle $[a_1, b_1]$ and $N_2$ D-branes around the three cycle
$[a_1, b_2]$ and so on $N_k$ D-branes around the three
cycle $[a_1, b_k]$ such that $N_1+N_2+ \cdots+ N_k=N_c$ 
and $N_f$ D-branes around the three  cycle $[b_1, a_2]$ and
$N_f$ D-branes around the three  cycle $[b_2, a_2]$ and so on
$N_f$ D-branes around the three  cycle $[b_k, a_2]$.
Therefore the final configuration after all the $b_i$'s are moved
to the left of $a_1$ in the configuration of points ordered as
$(a_1, b_1, \cdots, b_k, a_2)$ we get 
 is a configuration of points ordered
as $(b_1, b_2, \cdots, b_k, a_1, a_2)$ with $(N_f-N_1)$ D-branes wrapped around
$[b_1, b_2]$, $(2 N_f-N_1-N_2)$ D-branes wrapped around $[b_2, b_3]$
and so on
$((k-1) N_f-N_1-N_2- \cdots N_{k-1})$ D-branes around $[b_{k-1}, b_k]$ and
 $(k N_f-N_1-N_2- \cdots -N_k)$ D-branes wrapped around
$[b_k, a_1]$, $k N_f$ D-branes wrapped around $[a_1, a_2]$. In the limit 
$b_i (i= 1,2, \cdots, k-1) \rightarrow b_k$, the gauge group 
$SU(\widetilde{N_c})=
SU(k N_f-N_c)$ appears. In this case there are singlet fields, $M_i(i=1,
\cdots, k)$ coupled to the magnetic quarks.

{\bf 2) $SO(N_c)$ with a traceless symmetric tensor and $N_f$ vectors
($Sp(N_c)$ with a traceless antisymmetric tensor and
$N_f$ flavors) \cite{Intril,ILS}:}

In order to study  $SO(N_c)$  and $Sp(N_c)$ theories, as done in\cite{OV}
consider the local model of the CY threefold given by
$$
x^2+y^2= -\prod_i (z-a_i)(z-a'), \;\;\;\;  x'^2+y'^2=-z
$$
where $a_i$'s and $a'$ are real numbers with $a_1< a_2 <\cdots <a_k <0 < a'$.
Observe that the $S^2 \times S^1$ associated with $[a_{i-1}, a_{i}]$ 
for $i< k$ is realized either by real values for $x, y, x', y', z$ or
by purely imaginary values for $x, y, x', y'$ but real values for $z$. 
Also note  that the $S^3$ associated
with $[a_k, 0]$ is realized by taking the real
values of $x, y, x', y'$ and  $z$ while the $S^3$ associated
with $[0, a']$ is realized
by taking the imaginary values of $x, y, x'$ and $y'$  and real
value for $z$.

We discuss supersymmetric Yang-Mills theory with gauge group $SO(N_c)$
where the field $X$ is in the $\frac{N_c(N_c+1)}{2}-1$
traceless symmetric tensor representation 
of $SO(N_c)$ ($X_{ab}=X_{ba}$ and $X_{ab} \delta^{ab}=0$),
$N_f$ fields $Q^i$ are in the $N_c$ dimensional vector representation
of $SO(N_c)$ ($i=1, \cdots, N_f$). The superpotential is
$\mbox{Tr} X^{k+1}$. The theory which has this superpotential has
a stable vacuum provided $ N_f \geq \frac{N_c}{k}-4$.
For clear understanding let us first analyze the simplest case for the 
case of $k=2$(Of course, for $k=1$ case we have already
seen in\cite{OV} after integrating out massive $X$ and $Y$ 
that magnetic dual can be described in terms of the 
configuration of $(\frac{N_f}{2}-\frac{N_c}{2}+2)$ 
D-brane charge on the cycle 
$[0,a]$ and $\frac{N_f}{2}$ D-brane charge on $[a, a']$).
We wrap $N_1$ D-branes around $[a_1, 0]$ and $N_2$ D-branes around 
$[a_2, 0]$ and each of them has $N_f$ D-branes around $[0, a']$ such that
$N_1+N_2=N_c$ which can be understood that the number of D-branes on the
cycle $[a_2,0]$ should be $N_c$ as $a_1$ gets close to $a_2$.
Now the conjugation 
$$
(x,y,x', y' z) \mapsto (x^*, y^*, x'^*, y'^*, z^*),
$$ 
together with exchange of left- and right-movers on the world sheet,
will produce an orientifolding of the above configuration. The conjugation
preserves the above equation for $a_i$ and $a'$ real. In view of 
type I' theory,
these $D$-branes must be counted as $\frac{N_1}{2}$, $\frac{N_2}{2}$
and 
$\frac{N_f}{2}$ $D$-branes after orientifolding since the orientifolding
leaves these cycles invariant.
We get D-brane charge of the $[a_2, 0]$ cycle of $\frac{N_2}{2}\mp 2$.
Here the factor $\mp 2$ is  due to a contribution 
from the orientifold plane in addition to the physical D-branes,
the upper sign corresponds to the $SO(N_c)$ gauge group and
the lower one does the $Sp(N_c)$ gauge group. The
D-brane charge of the $[0, a']$ cycle is  $\frac{N_f}{2}$ after the action
of the orientifolding on the D-branes. By passing the point $a_2$ through
the point $0$ directly along the real axis due to the fact that
this operation should keep the orientifolding, 
the D-brane charge gets changed to the value of 
$\frac{N_f}{2}-\frac{N_2}{2}\pm 2$ on $[0, a_2]$ where
we assumed $N_f \geq N_2 \mp 4$. Next we move the point
$a_1$ to the positive real axis. The final configurations are given
by the $(-\frac{N_1}{2}\pm 2)$ D-brane charge on $[0, a_1]$, 
$(\frac{N_f}{2}-\frac{N_2}{2}\pm 2)$ D-brane charge
around $[0, a_2]$, $\frac{N_f}{2}$ D-brane charge on $[a_2, a']$ and
$\frac{N_f}{2}$ D-brane charge on $[0, a']$. Recombining the last 
$\frac{N_f}{2}$ D-branes into the D-branes on $[0, a_2]$ cycle and taking
the limit of $a_1 \rightarrow a_2$, it leads to 
$N_f-\frac{N_1}{2}-\frac{N_2}{2}+4$ that shows our expression for the magnetic
dual group for $k=2$ case,
$SO(\widetilde{N_c})= SO(2 N_f-N_c + 8)$
since there is no orientifold plane for $a_1, a_2 > 0$ and all these
D-brane charges are physical D-branes.
For the case of $Sp$ group with a traceless antisymmetric tensor and $N_f$
flavors, by recognizing that in the 
convention of
\cite{ILS} the symplectic group whose fundamental representation is 
$2N_c$ dimensional
as $Sp(N_c)$ and a flavor of it has two fields in the fundamental 
representation therefore $2 N_f$ fields
we obtain $2 \widetilde{N_c}=2(2N_f)-2 N_c -8$ which gives rise to
the following dual description $Sp(\widetilde{N_c})= Sp(2 N_f-N_c - 4)$.
 
For general value of $k$,
we get  
D-brane charges on the $[a_i, 0]$ cycle of $\frac{N_i}{2}\mp 2$ and
D-brane charges on the $[0, a']$ cycle of $\frac{N_f}{2}$ for
each $i=1, 2, \cdots, k$ after the action
of the orientifolding on the D-branes where we have $N_1+N_2+\cdots +N_k=N_c$.
The final configurations, after we moved all the $a_i$'s to the right of
the position of zero by successively doing similar things for the previous
case, are given
by the $(-\frac{N_1}{2}\pm 2)$ D-brane charge on $[0, a_1]$, 
the $(-\frac{N_2}{2}\pm 2)$ D-brane charge on $[0, a_2]$ and so on
the $(-\frac{N_{k-1}}{2}\pm 2)$ D-brane charge on $[0, a_{k-1}]$ and
$(\frac{k N_f}{2}-\frac{N_1+N_2+ \cdots +N_k}{2}\pm 2k)$ D-brane charge
around $[0, a_k]$, $\frac{k N_f}{2}$ D-brane charge on $[a_k, a']$.
In the limit of $a_i(i=1, 2, \cdots, k-1) \rightarrow a_k$,
we get the magnetic dual group, $SO(\widetilde{N_c})=SO(k( N_f+4)-N_c)$.
By similar reasoning for the case of $k=2$, we obtain the magnetic dual 
gauge group
for symplectic group as $Sp(\widetilde{N_c})=Sp((k-2) N_f-N_c)$.

{\bf 3) $SO(N_c)$ with an adjoint field and $N_f$ vectors
($Sp(N_c)$ with an adjoint field and $N_f$ vectors) \cite{LS,ILS}:}

We analyze supersymmetric Yang-Mills theory with gauge group $SO(N_c)$
where the adjoint field $X$ is in the $\frac{N_c(N_c-1)}{2}$
dimensional antisymmetric tensor  
of $SO(N_c)$ ($X_{ab}=-X_{ba}$),
$N_f$ flavors $Q^i$ are in the $N_c$ dimensional vector representation
of $SO(N_c)$ ($i=1, \cdots, N_f$). The superpotential is
$\mbox{Tr} X^{2(k+1)}$.
The theory given by this superpotential has a stable vacuum for
$N_f \geq \frac{N_c-4}{2 k+1} $.
For simplicity, we will start with the case of $k=1$.
Let us consider the case of wrapping $N_0$ D-branes around $[a, 0]$ and
$N_f$ D-branes around $[0, a']$. After orientifolding, the net D-brane
charge of $[a, 0]$ cycle becomes $\frac{N_0}{2}\mp 2$ and that of $[0, a']$
cycle is $\frac{N_f}{2}$. Now we bring other D-branes to the left
hand side of the point $a$ in the real axis whose D-brane charge of $[b_1, 0]$
cycle is $N_1$ and that of $[0, a']$ cycle is $N_f$.
In order to be consistent with the number of D-branes when we take the 
limit of $b \rightarrow a$, we should have $N_0+2 N_1=N_c$. 
We push the point $a$ along the real axis to the right and pass the point 
$0$.
In order to count the number of D-branes wrapped around $[0, a]$ and
$[a, a']$ we use the D-brane charge conservation and the orientation of
the D-branes. Then we have $(\frac{N_f}{2}-\frac{N_0}{2}\pm 2)$ D-brane charge
on $[0, a]$ and $\frac{N_f}{2}$ D-branes on $[a, a']$. 
Next we take the point $b$ along the real axis from negative to
positive values. The D-brane charge on $[0, b_1]$ is $-N_1$ due to the
orientation.
The $N_f$ D-branes which were going between $(0, a')$ can be decomposed
into $N_f$ D-branes between $(0, a)$ and $N_f$ D-branes between $(a, a')$.
Therefore the final picture we end up with is that there are
$-N_1$ D brane charge on $(0, b_1)$ and $(\frac{3 N_f}{2}-\frac{N_0}{2}\pm 2)$
on $(0, a)$ and $\frac{3 N_f}{2}$ on $(a, a')$. 
In the limit of $b_1 \rightarrow a$,
the magnetic dual group can be written as $SO(\widetilde{N_c})=
SO(3  N_f-N_c+4)$ by adding the two contributions. 
By twicing the $N_f$ and $N_c$ and dividing by two which leads to
$\frac{3(2N_f)-2N_c-4}{2}$,
we get $Sp(\widetilde{N_c})=Sp(3 N_f-N_c-2)$ for the symplectic group.

For the general value of $k \geq 2$, suppose we have the following picture:
after orientifolding, the net D-brane
charge of $[a, 0]$ cycle becomes $\frac{N_0}{2}\mp 2$ and that of $[0, a']$
cycle is $\frac{N_f}{2}$ and we bring other D-branes to the left
hand side of the point $a$ in the real axis whose 
D-brane charges of $[b_i, 0]$
cycle are $N_i$ and that of $[0, a']$ cycle are $N_f$ for each $i=1,
2, \cdots, k$.
In this case we also have the following relation, $N_0+2(N_1+
\cdots +N_k)=N_c$.
Then the final configuration is that there are
$-N_i$ D brane charges on $(0, b_i)$ for each $i=1, 2, \cdots, k$
 and $\frac{(2k+1) N_f}{2}-\frac{N_0}{2}\pm 2-N_1-N_2- \cdots -N_k$
on $(0, a)$ and $\frac{(2k+1) N_f}{2}$ on $(a, a')$. 
In the limit of $b_i(i=1, 2,
\cdots, k) \rightarrow a$,
the dual theory has gauge group, $SO(\widetilde{N_c})=
SO((2 k+1) N_f-N_c+4)$ indicating $(2 k+1)$ mesons coupled dual quarks and 
$ Sp(\widetilde{N_c})=Sp((2 k+1) N_f-N_c-2)$.

{\bf 4) $SU(N_c)$ with a symmetric flavor and $N_f$ fundamental
flavors
($SU(N_c)$ with an antisymmetric flavor and $N_f$ fundamental flavors)
\cite{ILS}:}

The fields $X$ and $\widetilde{X}$ are a flavor of symmetric tensor
representations of $SU(N_c)$ and there are   
$N_f$ fundamental multiplets $Q^i$, and $N_f$ antifundamental
multiplets $\widetilde{Q}_{\widetilde{i}}$ where $i,
\widetilde{i}=1, \cdots, N_f$.
The superpotential is $\mbox{Tr} (X \widetilde{X})^{k+1}$.
Now we continue to repeat the procedure we have done so far for the
case of ordered configuration as $(b, a_1, a_2, 0, a')$ when we
consider $k=2$ case (Recall that when $k=1$, the configuration
was $-N_0$ D-brane charge on $[0,b]$ and $ N_f-\frac{N_1}{2} 
\pm 2$ on $[0, a_1]$
and $N_f$ on $[0, a']$). 
After orientifolding,  D-brane charges are $N_0$ around 
$[b, 0]$, $\frac{N_1}{2}\mp 2$ around $[a_1, 0]$, $\frac{N_2}{2}\mp 2$ around
$[a_2, 0]$. Each of them produces $N_f$ D-brane charge around
 $[0, a']$ where $N_0+N_1+N_2=N_c$.
We have seen already ordered configuration as $(a_1, a_2, 0, a')$ in the 
previous subsection.
The final configuration is given
by the $(-\frac{N_1}{2}\pm 2)$ D-brane charge on $[0, a_1]$, 
$(N_f-\frac{N_2}{2}\pm 2)$ D-brane charge
around $[0, a_2]$, $N_f$ D-brane charge on $[a_2, a']$ and
$N_f$ D-brane charge on $[0, a']$. 
Next we move the point $b$ along the real axis from negative to
positive values. The D-brane charge on $[0, b]$ is $-N_0$ due to the
orientation. 
Therefore the final configuration is given
by the $-N_0$ D-brane charge on $[0, b]$,
the $(-\frac{N_1}{2}\pm 2)$ D-brane charge on $[0, a_1]$, 
$(N_f-\frac{N_2}{2}\pm 2)$ D-brane charge
around $[0, a_2]$, $N_f$ D-brane charge on $[a_2, a']$ and
$2 N_f$ D-brane charges on $[0, a']$. Recombining one of $N_f$'s
D-branes into the D-branes on $[0, a_2]$ cycle and taking
the limit of $a_1 \rightarrow a_2$, we obtain 
$2 N_f-\frac{N_1}{2}-\frac{N_2}{2} \pm 4$. Finally we see that
the dual theory has the gauge group $SU(\widetilde{N_c})=SU(5 N_f-N_c+8)$
by taking into account of the contributions from
$-N_0$ D-brane charge on $[0, b]$ and $N_f$ on $[0, a']$.
On the other hand, when we consider antisymmetric flavor 
instead of symmetric one
we get $SU(\widetilde{N_c})=SU(5 N_f-N_c-8)$.

For the general value of $k$, after orientifolding 
D-brane charges are $N_0$ around 
$[b, 0]$, 
$\frac{N_i}{2}\mp 2$ around $[a_i, 0]$ for each $i=1, 2, \cdots, k$
and each of them has $N_f$ D-brane charges around $[0, a']$ respectively 
where
$N_0+N_1+N_2+\cdots + N_k=N_c$.
Since $2(k N_f-\frac{1}{2}(N_1+ \cdots +N_k) \pm 2 k)-N_0+N_f=(2 k+1)N_f-N_c
\pm 4 k$, we arrive at the magnetic dual group for $SU$,
$SU(\widetilde{N_c})=SU((2 k+1)N_f+4 k-N_c)$ and that for $SU(N_c)$ with
an antisymmetric flavor and $N_f$ fundamental flavors is 
$SU(\widetilde{N_c})=SU((2 k+1)N_f-4 k-N_c)$.

{\bf 
5) $SU(N_c)$ with an antisymmetric tensor and a symmetric tensor \cite{ILS}:}

In this case the field $X$ is in the $\frac{N_c(N_c-1)}{2}$ representation,
the field $\widetilde{X}$ in the $\overline{\frac{N_c(N_c+1)}{2}}$
representation, $m_f(\widetilde{m_f})$ fields                            
 $Q^i(\widetilde{Q_{\widetilde{i}}})$ in the
(anti)fundamental representation. The superpotential is given by
$\mbox{Tr}(X\widetilde{X})^{2(k+1)}$.
We expect that the above procedure of case 1)
can be applied similarly, for example, $k=1$. 
We wrap $N_0$ D-branes around the
three cycle $[a_1, b_0]$ and $2 N_1$ D-branes around the three cycle
$[a_1, b_1]$ such that $N_0+2 N_1=N_c$ 
and $\frac{3(m_f+\widetilde{m_f})}{2}$ D-branes around the 
 cycle $[b_0, a_2]$ and $2(m_f+\widetilde{m_f})$
D-branes around the cycle $[b_1, a_2]$.
The final configuration after we move $b_0$ and $b_1$ to the left
of $a_1$ in the configuration of point ordered as $(a_1, b_0, b_1, a_2)$
we get 
is a configuration of points ordered
as $(b_0, b_1, a_1, a_2)$ with $(\frac{3(m_f+\widetilde{m_f})}{2}-N_0)$ 
D-branes wrapped around
$[b_0, b_1]$, $(\frac{7(m_f+\widetilde{m_f})}{2}-N_0-2 N_1)$ 
D-branes wrapped around $[b_1, a_1]$,
$\frac{7(m_f+\widetilde{m_f})}{2}$
D-branes wrapped around $[a_1, a_2]$. In the limit 
$b_0 \rightarrow b_1$, the gauge group 
becomes $SU(\widetilde{N_c})=
SU(\frac{7(m_f+\widetilde{m_f})}{2}-N_c)$. 

For general value of $k$, we wrap $N_0$ D-branes around the cycle 
$[a_1, b_0]$ and 
$2 N_i$ D-branes around the 
cycle $[a_1, b_i]$ for each $i=1, 2, \cdots, k$ 
such that $N_0+2 N_1+ \cdots+2 N_k=N_c$ 
and $\frac{3(m_f+\widetilde{m_f})}{2}$ D-branes 
around the cycle $[b_0, a_2]$ and
$2(m_f+\widetilde{m_f})$ D-branes around the cycle $[b_i, a_2]$ for each $i$.
The final configuration 
 is a configuration of points ordered
as $(b_0, b_1, \cdots, b_k, a_1, a_2)$ with 
$(\frac{3(m_f+\widetilde{m_f})}{2}-N_0)$ D-branes wrapped around
$[b_0, b_1]$, $(\frac{7(m_f+\widetilde{m_f})}{2}-N_0-2 N_1)$ 
D-branes wrapped around $[b_1, b_2], \cdots,
 ((2k+\frac{3}{2})(m_f+\widetilde{m_f})-N_0-2 N_1- \cdots -
2 N_k)$ D-branes wrapped around
$[b_k, a_1]$, and $(2 k+\frac{3}{2})(m_f+\widetilde{m_f})$ 
D-branes wrapped around $[a_1, a_2]$. In the limit 
$b_i (i= 0, 1, \cdots, k-1) \rightarrow b_k$, the gauge group 
$SU(\widetilde{N_c})=
SU((2 k+\frac{3}{2})(m_f+\widetilde{m_f})-N_c)$ appears. 


The authors thank Dept. of Mathematics,
Pohang University of Science and Technology
for the hospitality where this work has been done.


\vskip -1cm
\noindent
\begin{thebibliography}{[00]}
\baselineskip 20pt
\bibitem{KV1} S. Katz and C. Vafa, {\it Geometric Engineering of $N=1$
Quantum Field Theories}, hep-th/9611090.
\bibitem{BJPSV} M. Bershadsky, A. Johansen, T. Pantev, V. Sadov and 
C. Vafa, {\it F-theory, Geometric Engineering and $N=1$ 
Dualities}, hep-th/9612052.
\bibitem{Seiberg} N. Seiberg, {\it Electric Magnetic Duality in
Supersymmetric Non-Abelian Gauge Theories}, hep-th/9411149.
\bibitem{VZ} C. Vafa and B. Zwiebach, {\it $N=1$ Dualities of $SO$ and 
$USp$ Gauge Theories and T-Duality of String Theory}, hep-th/9701015.
\bibitem{IS} K. Intriligator and N. Seiberg, {\it Duality, Monopoles,
Dyons, Confinement and Oblique Confinement in Supersymmetric 
$SO(N_c)$ Gauge Theories}, hep-th/9503179.
\bibitem{IP} K. Intriligator and P. Pouliot, {\it Exact Superpotentials,
Quantum Vacua and Duality in Supersymmetric $SP(N_c)$ Gauge 
Theories}, hep-th/9505006.
\bibitem{EGK} S. Elitzur, A. Giveon and D. Kutasov, {\it Branes and
$N=1$ Duality in String Theory}, hep-th/9702014.
\bibitem{HW} A. Hanany and E. Witten, {\it Type IIB Superstrings,
BPS Monopoles, and Three-Dimensional Gauge Dynamics}, hep-th/9611230.
\bibitem{EJS} N. Evans, C. Johnson and A. Shapere, {\it Orientifolds,
Branes, and Duality of 4D Gauge Theories}, hep-th/9703210.
\bibitem{BH} J.H. Brodie and A. Hanany, {\it Type IIA Superstrings,
Chiral Symmetry, and $N=1$ 4D Gauge Theory Dualities}, hep-th/9704043.
\bibitem{BSTY} A. Brandhuber, J. Sonnenschein, S. Theisen and
S. Yankielowicz, {\it Brane Configuration and 4D Field Theory Dualities},
hep-th/9704044.
\bibitem{OV} H. Ooguri and C. Vafa, {\it Geometry of $N=1$ Dualities in
Four Dimensions}, hep-th/9702180.
\bibitem{Kutasov} D. Kutasov,
 {\it A Comment on Duality in $N=1$ Supersymmetric
Non-Abelian Gauge Theories}, hep-th/9503086; D. Kutasov and A. Schwimmer,
{\it On Duality in Supersymmetric Yang-Mills Theory}, hep-th/9505004.
\bibitem{Intril} K. Intriligator, {\it New RG Fixed Points and Duality in
Supersymmetric $SP(N_c)$ and $SO(N_c)$ Gauge Theories}, hep-th/9505051.
\bibitem{LS} R.G. Leigh and M.J. Strassler, {\it Duality of $Sp(2 N_c)$ and
$SO(N_c)$ Supersymmetric Gauge Theories with Adjoint Matter}, 
hep-th/9505088.
\bibitem{ILS} K. Intriligator, R.G. Leigh and M.J. Strassler, 
{\it New Examples of Duality in Chiral and Non-Chiral Supersymmetric
Gauge Theories}, hep-th/9506148.
\bibitem{CDFV} A. Ceresole, R. D'Auria, S. Ferrara and A. Van Proeyen,
{\it Duality Transformations in Supersymmetric Yang-Mills Theories
Coupled to Supergravity},  hep-th/9502072.
\bibitem{BSV} M. Bershadsky, V. Sadov and C. Vafa, 
{\it D-Strings on D-Manifolds}, hep-th/9510225.
\end{thebibliography}
\end{document}